\documentclass[fleqn,usenatbib]{mnras}  

\usepackage{newtxtext,newtxmath}
\usepackage[T1]{fontenc}
\usepackage{ae,aecompl}

\usepackage{graphicx}	% Including figure files
\usepackage{amsmath}	% Advanced maths commands
\usepackage{amssymb}	% Extra maths symbols
\usepackage{multirow}   %multiple rows in tables

\title[OMEGA V: Jellyfish Galaxies in A901/2]{OMEGA -- OSIRIS Mapping of Emission-line Galaxies in A901/2 -- V. The rich population of jellyfish galaxies in the multi-cluster system Abell 901/2}

\author[F. V. Roman-Oliveira et al.]{
\parbox{\textwidth}{Fernanda V. Roman-Oliveira$^{1}$,
Ana L. Chies-Santos$^{1}$,
Bruno Rodr\'iguez del Pino$^{2}$,
A. Arag\'on-Salamanca$^{3}$,
Meghan E. Gray$^{3}$,
Steven P. Bamford$^{3}$}
\vspace{0.3cm}\\
$^{1}$Departamento de Astronomia, Instituto de F\'isica, Universidade Federal do Rio Grande do Sul, Porto Alegre, RS, Brazil\\
$^{2}$Centro de Astrobiolog\'ia, CSIC-INTA, Torrej\'on de Ardoz, 28850, Madrid, Spain \\
$^{3}$School of Physics and Astronomy, The University of Nottingham, University Park, Nottingham NG7 2RD, UK \\
}

\date{Accepted 2018 December 10. Received 2018 November 23; in original form 2018 October 12}

\pubyear{2018}

\begin{document}
\label{firstpage}
\pagerange{\pageref{firstpage}--\pageref{lastpage}}
\maketitle

\begin{abstract}%
We present the results of a systematic search and characterisation of galaxies with morphological signatures of ram-pressure stripping, known as jellyfish galaxies, in the multi-cluster system A901/2, at z $\sim$ 0.165, as part of the OMEGA survey. By visual inspecting ACS/HST F606W images looking for morphological signatures of ram-pressure stripping events in H$\alpha$-emitting galaxies, we identify a total of 70 jellyfish candidates. Out of these, 53 are clearly star-forming galaxies and 5 are highly probable AGN hosts, the classification of the remaining galaxies is more uncertain. 
They have late-type and irregular morphologies and most of them are part of the blue cloud with only 4 being previously classified as dusty reds. The AGN activity is not prominent in the sample and, of the few cases of galaxies hosting AGN, such activity does not seem to be correlated to the gas stripping phenomenon. Our jellyfish galaxy candidates do not have a preferential pattern of motion within the multi-cluster system, although the most compelling cases appear to inhabit the inner regions of the most massive sub-cluster centres. The specific star-formation rate of these galaxies indicates that their star formation activity is enhanced, in contrast with what is observed for the rest of the star-forming galaxy population in the system. Half of the sample is forming stars at a higher rate than the main-sequence for field galaxies and this behaviour is more evident for the most compelling candidates. For some galaxies, the spatially resolved H$\alpha$ emission appears to be as disturbed and extended as their continuum counterparts. Our findings point towards a scenario where the ram pressure stripping is triggering a period of intense and extended star formation throughout the galaxy while it is also disturbing the morphology. This is the largest sample of jellyfish galaxy candidates found in a single system suggesting that cluster mergers might be the ideal environment for studying ram pressure stripping effects.
\end{abstract}


%%%%%%%%%%%%%%%%%%%%%%%%%%%%%%%%%%%%%%%%%%%%%%%%%%

%%%%%%%%%%%%%%%%% BODY OF PAPER %%%%%%%%%%%%%%%%%%

\section{Introduction}
The environment in which galaxies inhabit influences their physical properties and evolution.
As they interact with their surroundings, their morphologies and star formation properties can be severely changed.
The low presence of early-type galaxies in the field and its dominance in denser regions of the Universe points towards a scenario in which environmental mechanisms play a major role in galaxy quenching and morphological evolution \citep{dressler80}.

Such transformations can be driven both by internal properties and processes, e.g. mass \citep[]{baldry06}, supernovae and AGN feedback \citep{newtonkay13, boothschaye09}; and external ones, such as tidal interactions or mergers \citep{barnes92}, galaxy harassment \citep{moore96} and ram pressure stripping \citep{gunngott72}; the latter being more common in high density environments.
Although there are several physical mechanisms competing, the dominance and extent of each one are not yet fully comprehended.
For this reason, the morphological and physical changes that the environment induces in galaxies is crucial to the understanding of galaxy evolution as a whole.

Ram pressure stripping (RPS) is the interaction that occurs when a galaxy rich in gas falls toward a denser region, such as the core of a galaxy cluster, and it experiences the stripping of its cold gas as a result of a hydrodynamical friction with the hot and dense intracluster medium (ICM) \citep{gunngott72}.
It is cited as one of the most efficient mechanisms in quenching star formation in clusters \citep{boselli16, simpson18}, but it has also been suggested that, for a short period of time, it could enhance the star formation due to turbulences in the galaxy causing cold gas clouds to collapse \citep{bekkicouch03}.
Galaxies undergoing RPS also tend to display intense star formation in their outskirts, in the shape of severely disturbed debris holding clumps of young stars \citep{cortese07, yagi10, rawle14, fumagalli14, ebeling14, mcpartland16}.
The loss of the gas reservoir of a galaxy undergoing RPS can soon lead to a more passive existence, linking such process to the quenching of star formation in galaxies rich in gas in cluster environments \citep{jaffe16, vollmer12}.
However, it is not always the case that the star formation is found to be enhanced.
Some hydrodynamical simulations suggest that the quenching or enhancement could be a factor of galaxy properties, such as the inclination of the disk during the infalling on the cluster \citep{bekki14, steinhauser16}.

In the most extreme cases of galaxies undergoing RPS, the debris and gas trails can conglomerate unilaterally and extend to the opposite direction of motion.
These cases can transform the morphology of the original galaxy in a way that resembles jellyfish-like creatures, hence their names. 
To our knowledge, the term jellyfish-like structure was first introduced by \citet{Bekki09}.
These galaxies have previously been found in low numbers in cluster environments (21 in Coma, \citealt{smith10}, \citealt{yagi10}, \citealt{gavazzi18}; 6 in Virgo, \citealt{abramson16}, \citealt{kenney14}, \citealt{kenney99}, \citealt{boselli16}, \citealt{boselli18}, \citealt{fossati18}; 2 in A3627, \citealt{sun06}, \citealt{sun07}, \citet{sun10}, \citealt{zhang13}; 1 in A1367, \citealt{yagi17}; 5 in A2744, \citealt{rawle14}).
Systematic searches for jellyfish galaxies in several different systems have also been carried out, most notably in the MACS (The MAssive Cluster Survey) clusters (z= 0.30-0.43) by \citet{ebeling14} and \citet{mcpartland16} as well as in the OMEGAWINGS+WINGS clusters (z= 0.04-0.07) by \citet{poggianti16}; the latter lead to the GASP (GAs Stripping Phenomena in galaxies with MUSE) survey, a large ESO/MUSE study on the ram pressure stripping phenomena \citep{poggianti17}.
Recently, jellyfish galaxies have been identified in the Illustris TNG simulations \citep{yun18}.

%tadpole comment
At first sight, the jellyfish morphology appears to resemble that of tadpole galaxies, objects first found in the higher redshift Universe probed by the Hubble Deep Field (HDF) \citep{vandenbergh96} and later studied in more detail in the Hubble Ultra Deep Field (HUDF) \citep{elmegreen07, elmegreen10, straughn15}.
These are galaxies with a diffuse tail attached to a head of a bright decentralised clumpy star-forming structure \citep{sanchezalmeida13}. However, the formation of tadpole galaxies cannot to be described entirely by the RPS phenomenon and there are numerous alternative proposed origins (see e.g. \citealt{sanchezalmeida13}).
A striking difference between jellyfish and tadpole galaxies is that the former present enhanced star formation in the tails region \citep{poggianti18}, while in the latter, star formation is enhanced in the head region \citep{vandenbergh96, abraham96}. It is also important to stress that the jellyfish phenomenon is associated with cluster environments, crucial to explain their origin, while that is not the case for tadpole galaxies.

%at such an early universe the largest structures are proto-clusters, therefore the origins of both these classes of objects are distinct. e.g. being the result of a merger or galaxy encounters \citep{}, disks with unilaterally triggered star formation \citep{sanchezalmeida13} or ... and subsequently by the While jellyfish galaxies are formed through the interaction with the ICM in massive clusters \citep{gunngott72}, t

The OMEGA survey was designed to generate deep, low-resolution spectra around the H$\alpha$ ($\lambda=$6563 \AA) and [NII] ($\lambda=$6548 \AA, $\lambda=$6584 \AA) emission-lines for all the galaxies in the Abell 901/2 multi-cluster system. This was accomplished with observations with the tunable-filter instrument OSIRIS located at the 10.4m Gran Telescopio Canarias (GTC).
The main goal of the OMEGA survey is to provide a better understanding on star formation and AGN activity across the A901/2 system by targeting the emission lines H$\alpha$ and [NII] in the whole area of the system \citep{chies-santos15, rodriguezdelpino17, weinzirl17, wolf18}.
The A901/2 system, at z $\sim$ 0.165, covers a 0.51 $\times$ 0.42 square degree area in the sky and its large range of different environments provides a great laboratory for galaxy evolution.
It has been observed in many wavelengths and extensively studied by the STAGES \citep{gray09} and COMBO-17 surveys \citep{wolf03}.
Moreover, the system has been observed with XMM-Newton, GALEX, HST, Spitzer, VLT/VIMOS, PRIMUS, 2dF and GMRT.

Constraining the properties of jellyfish galaxies is crucial to understand the role of the RPS phenomena in the environmental quenching we observe in galaxies in dense environments.
The combination of HST imaging \citep{gray09} and H$\alpha$ maps from the OMEGA survey is ideal to search for jellyfish galaxies and to study how this effect can alter the evolutionary path of galaxies in the environments probed in A901/2.
This paper is organised as follows: in Section \ref{sec:data} we describe the data used throughout the study; in Section \ref{sec:sample} we discuss the criteria used for selecting the sample of jellyfish galaxy candidates; in Section \ref{sec:res} we show and discuss the main results of our study by exploring their general properties, e.g. morphology, mass and SED types, as well as their star formation properties and spatial distribution as a function of environment; in Section \ref{conc} we present a summary of our findings and the conclusions.

Thorough the paper we adopt a $H_0= 70 km s^{-1} Mpc^{-1}$, $\Omega_\Delta=0.7$ and $\Omega_M=0.3$ cosmology.

\section{Data}\label{sec:data}

\subsection{The OMEGA Survey}
In this work we have used the integrated star formation rates and AGN/SF emission-line diagnostics from \citet{rodriguezdelpino17}. We have also used the H$\alpha$ spatially resolved emission stamps from Rodr\'iguez del Pino et al. in prep..

For a detailed description of the survey details, the data acquisition and reduction see \citet{chies-santos15}.
The analysis of the integrated star formation and AGN properties of the whole survey can be found in \citet{rodriguezdelpino17}. \citet{weinzirl17} performs the study of the phase-space properties of the OMEGA galaxies.
The study of how inclination affects different star formation estimators was done in \citet{wolf18}.

\subsection{Additional data for the Abell901/2 multicluster system}
In addition to the data from OMEGA, we have also used the ACS/WFC3 F606W Hubble Space Telescope (HST)/Advanced Camera for Surveys (ACS) images available from STAGES \citep{gray09}.
We have used some of the galaxy properties available in the STAGES catalogue \citep{gray09}, such as stellar masses, the SED types classification, previously visually assigned morphologies and stellar environmental densities.
The A901/2 galaxies were classified in three different SED types: blue cloud, old red and dusty red \citep{wolf05}. The blue cloud are blue normal star-forming galaxies while the old red are red passive galaxies. The dusty red have obscured star formation and have been shown to host active star formation on average four times lower than the blue cloud galaxies \citep{wolf09}. The term "dusty" may be misleading, as these galaxies do not have more dust than the other star-forming galaxies. As they have relatively low star formation, the same amount of dust makes them look redder.
Moreover, we have used the XMM-Newton X-ray image of the system for mapping the hot gas in the system \citep{gilmour07}.
We have also used the stamps from the RGB COMBO-17 poster for display purposes. These images are illustrative and have the sole purpose to provide a better view of the galaxies.

\section{The Sample}\label{sec:sample}%ok
\subsection{Sample Selection}%ok
In order to obtain a sample of jellyfish galaxy candidates in A901/2, we performed a search within the OMEGA sample of detected H$\alpha$-emitting sources from \citet{chies-santos15}, and \citet{rodriguezdelpino17}.
The OMEGA sample contains 439 H$\alpha$-emitting galaxies with masses ranging from 10$^{9}$ to 10$^{11.5}$ M$_\odot$ that are classified as members of the A901/2 system \citep{gray09}.
These galaxies can have active star formation and/or host AGN activity.
Given that jellyfish galaxies have been found to strongly emit in H$\alpha$ \citep{smith10, vulcani16, abramson16, bellhouse17, sheen17} it is a reasonable starting point to search for them in OMEGA.
Three of us (ACS, BRP and FRO) visually inspected the HST/F606W images searching for visual morphological features of gas stripping. Our classification scheme was based on the methods described in~\citet{ebeling14} and ~\citet{poggianti16}.

The visual inspection was first performed independently by each classifier who evaluated the presence of three main morphological features following \citet{ebeling14}:

\begin{enumerate}
    \item unilaterally disturbed morphology;
    \item bright knots of star formation;
    \item debris trails.
\end{enumerate}

According to the level of visual evidence of morphological features of stripping, each classifier assigned a JClass for each galaxy ranging from 0 to 5, following the method described in \citet{poggianti16}.
Starting from JClass 1 for the weakest evidences, the stronger cases were classified with higher JClasses up to the most extreme JClass 5 events.
Galaxies with no evidence of stripping were assigned JClass 0.
The JClasses 1 and 2 are galaxies that may show some weak visual evidence of stripping, but the evidence is not strong enough for selecting them as secure candidates.
The JClass 3 are galaxies with light visual evidences of stripping that are probable cases of galaxies undergoing a stripping event.
Finally, JClass 4 and 5 cover the strongest candidates.

We leave the weakest cases (JClasses 1 and 2) out of the final sample of jellyfish candidates as their physical origin is difficult to evaluate based solely on the images observed.

\begin{figure*}
\centering
\includegraphics{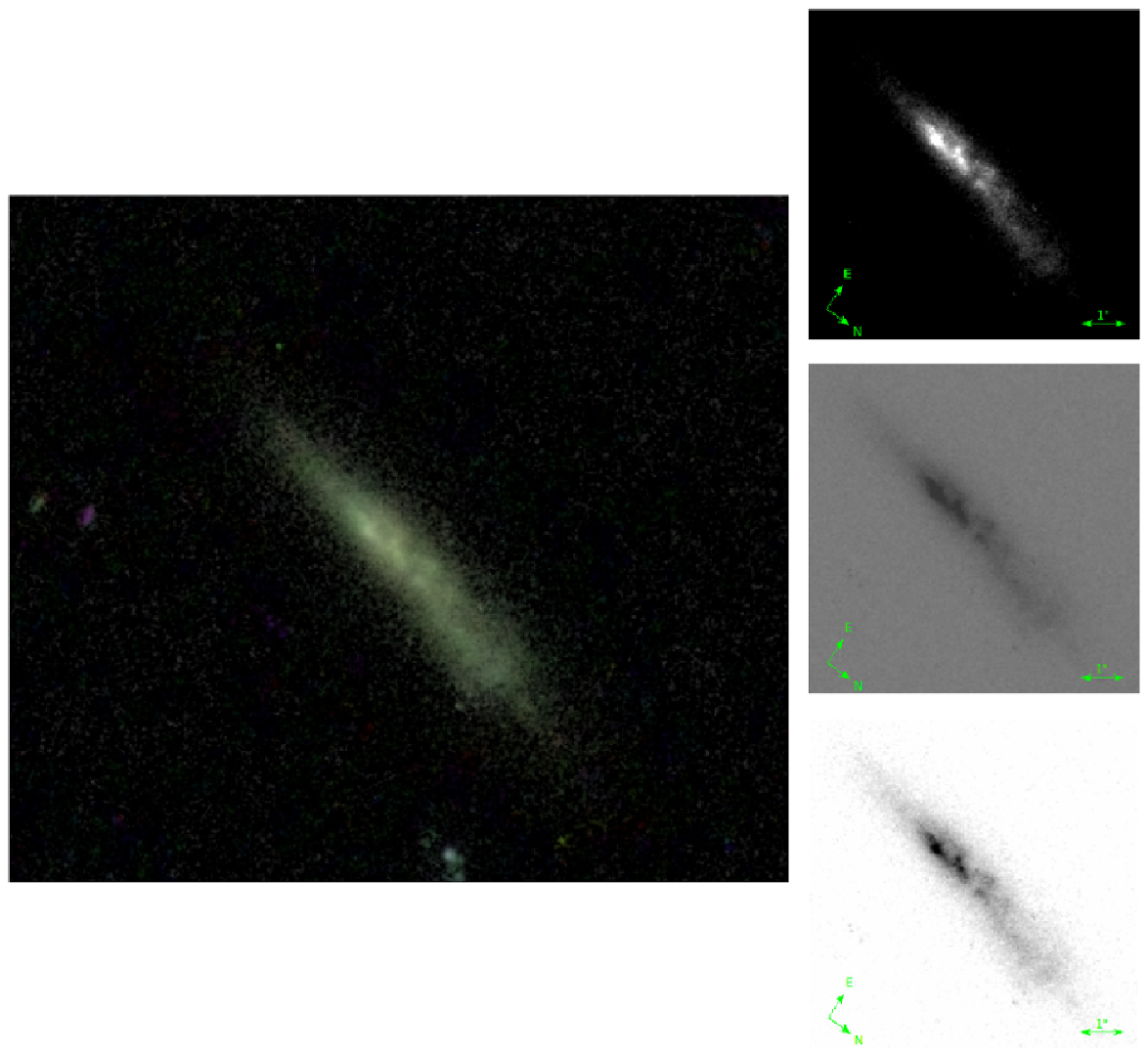}
\caption{Examples of jellyfish galaxy candidates, the upper panel shows a JClass 5, the strongest case, and each following panel shows the next consecutive lower JClass until reaching JClass 1, the weakest case, at the bottom panel. For each galaxy: on the left the composed RGB image from the COMBO-17 poster. On the right three different contrasts of the HST image allowing the observer to recognise the debris trails and knots.}
\label{fig:ex}
\end{figure*}

Our final sample of jellyfish candidates is selected by including those galaxies classified as JClass 3 or higher by at least two classifiers. We assign them a final JClass determined as the median of the three classifications. 
The final sample consists of 73 galaxies of which 11 galaxies are assigned a final JClass 5, 24 galaxies a final JClass 4 and the remaining 38 galaxies are assigned a final JClass 3.
The whole sample of candidates is presented in the ATLAS that is available online as a supplementary material to this article. The Figure ~\ref{fig:ex} shows one example of each of the JClass categories -- top to bottom panels: JClass 5, 4, 3, 2 and 1.

%%%%%%%%%%%%%%%%%%%EXAMPLES%%%%%%%%%%%%%%%%%%%

\begin{figure*} %45301 - JClass5
\centering
\includegraphics{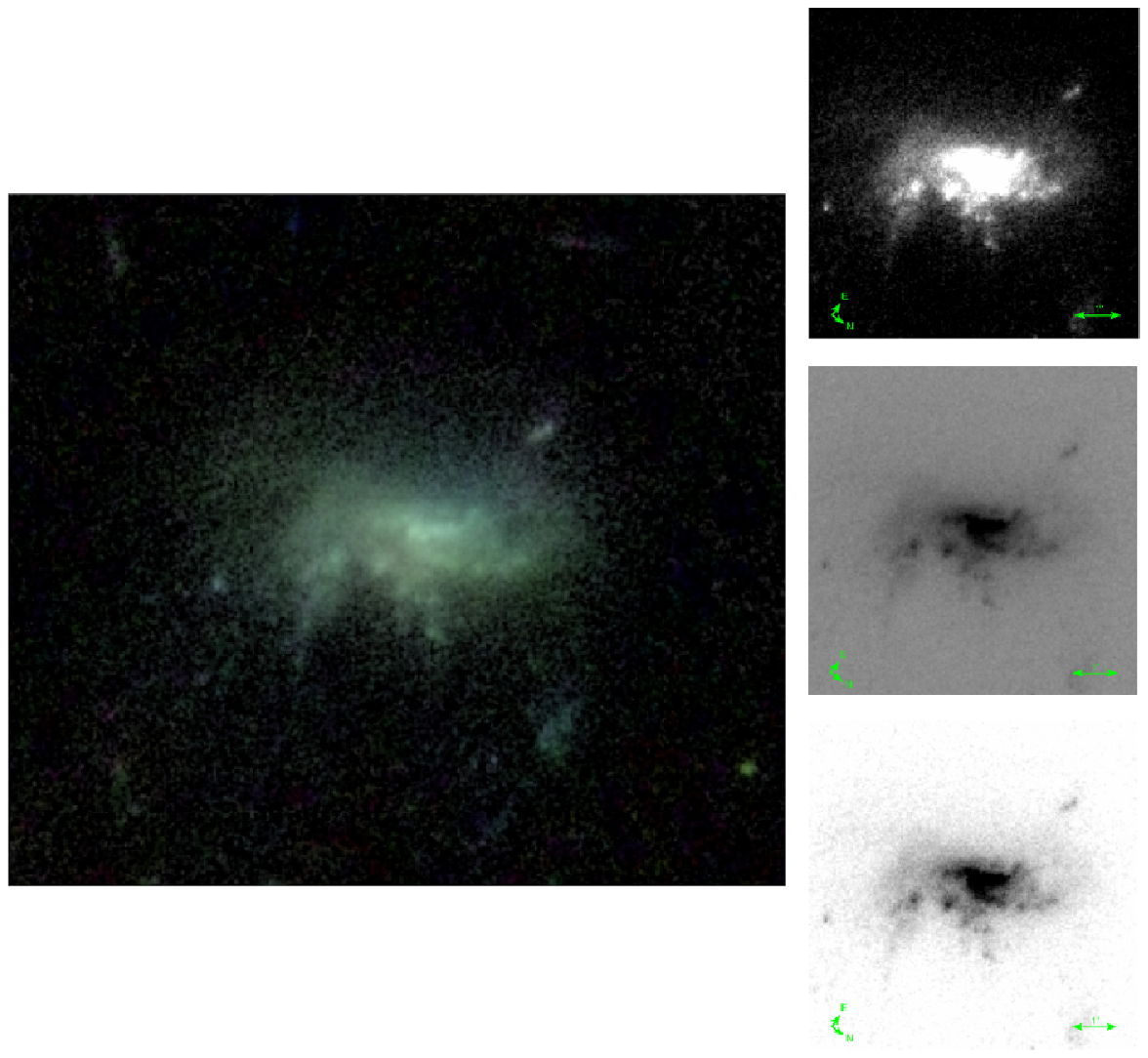}  
\end{figure*}

%20056 - JClass4
\begin{figure*}
\centering
\includegraphics{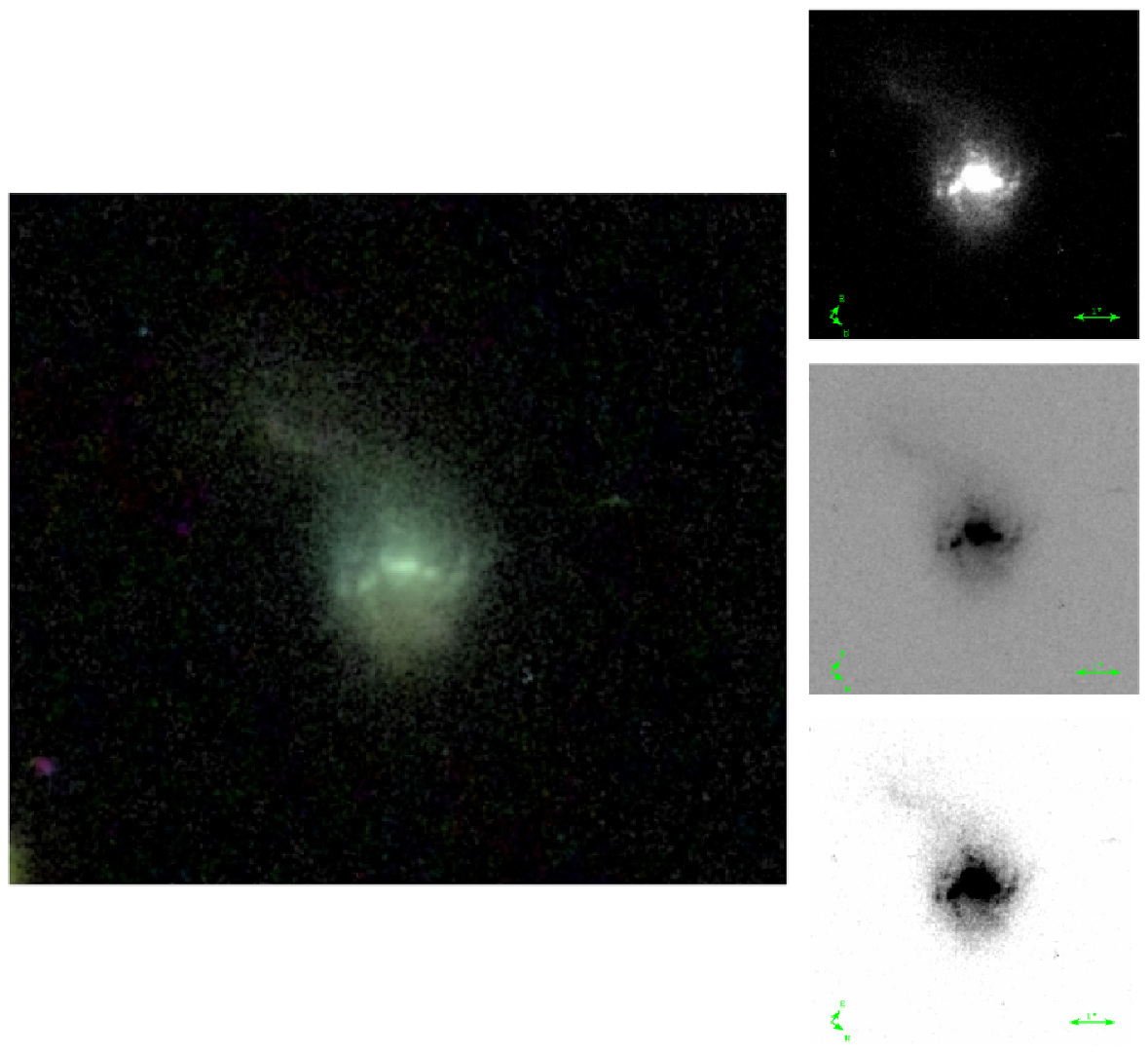}
\end{figure*}

%55188 - JClass 3
\begin{figure*}
\centering
\includegraphics{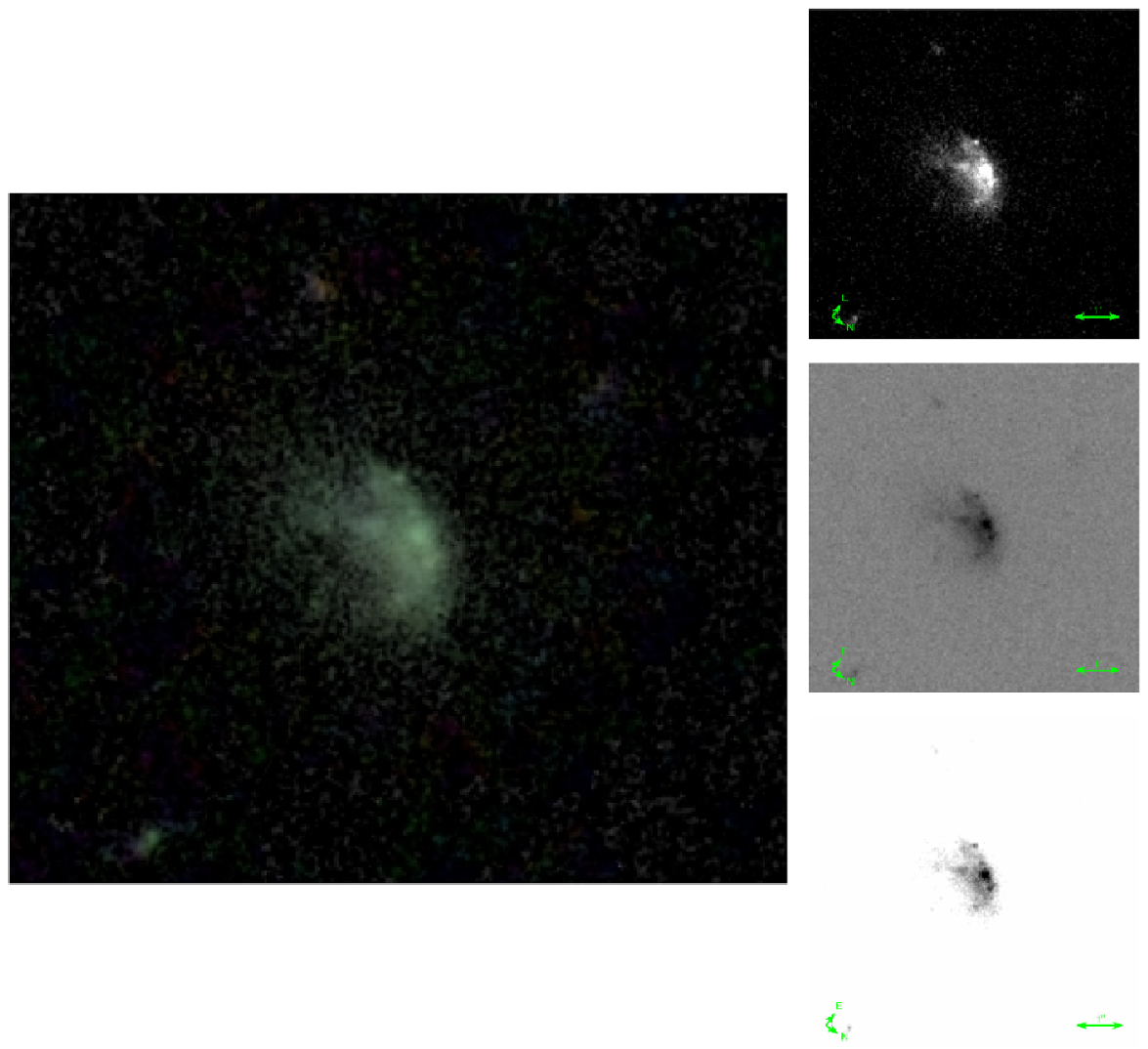}
\end{figure*}

%14975 - JClass 2
\begin{figure*}
\centering
\includegraphics{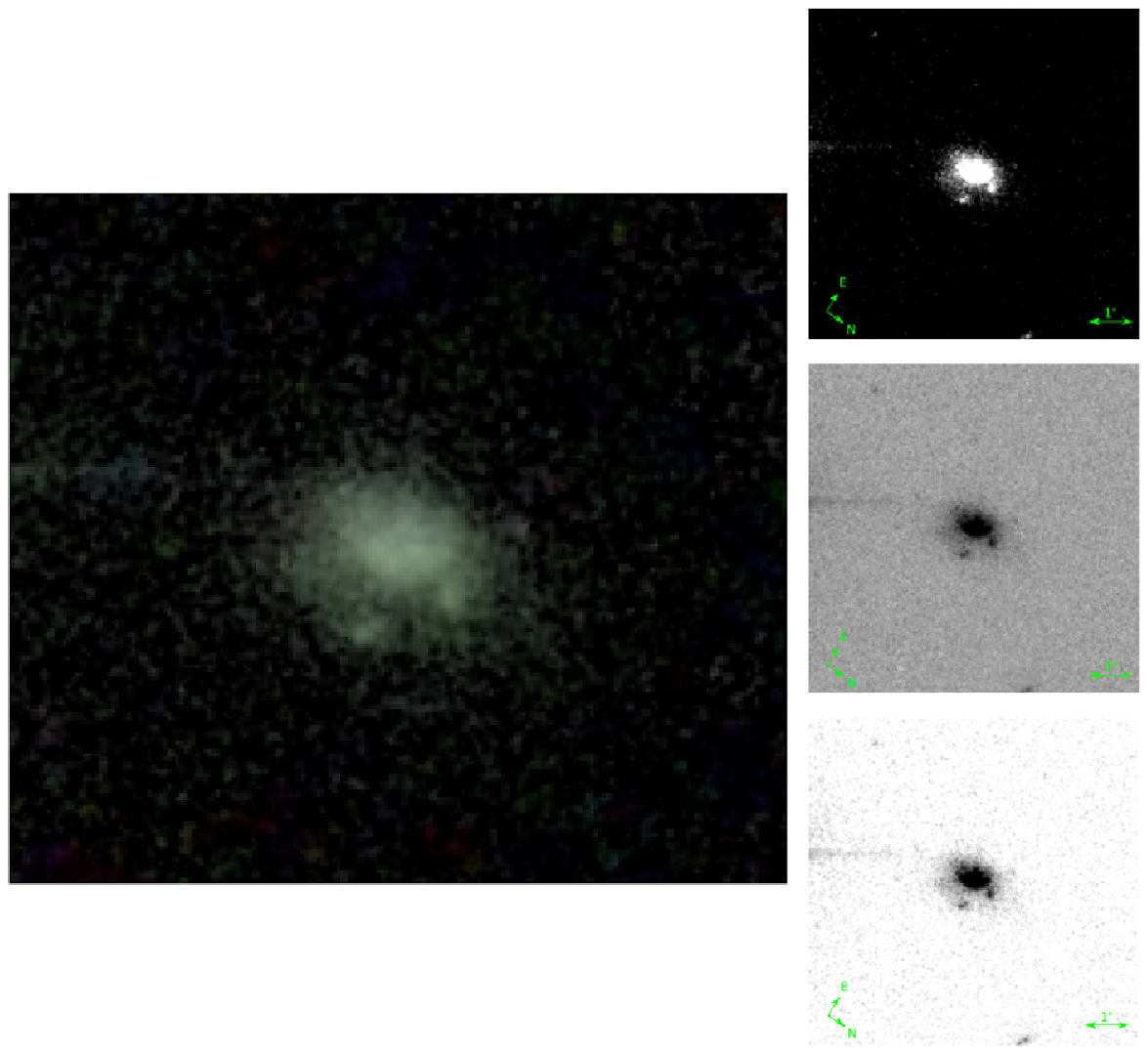}
\end{figure*}

%23457 - JClass 1

%%%%%%%%%%%%%%%%%%%EXAMPLES%%%%%%%%%%%%%%%%%%%

To verify whether the sample selection is biased because of using only H$\alpha$-emitting galaxies, we applied the same selection method to a control sample.
This control sample was composed by 200 random non-H$\alpha$-emitting galaxies that are confirmed cluster members and occupy the same range of mass. From the 200 galaxies, we found only one case of a JClass 4 and two cases of JClass 3.
Therefore, there is little morphological evidence of RPS in the control sample. This indicates that we are selecting the majority of jellyfish candidates with very low incompleteness in our H$\alpha$ detected sample. It also tightens the link between jellyfish galaxies and H$\alpha$ emission which is an indicator of recent star formation.

The selected sample can also be contaminated by galaxies that have irregular jellyfish-like morphologies because of other mechanisms non-related to RPS. These contaminants should be mainly galaxies that went through tidal interactions with close companions or mergers. For testing our sample for such contaminants, we have checked if the jellyfish galaxy candidates appear to be systematically closer to their neighbours than the other galaxies in the system. Measuring the projected distance to the closest neighbour for both the final jellyfish candidates sample and for a control sample of 450 random cluster members in the same range of mass, through a Kolmogorov-Smirnov (KS) test we find no significant difference between both populations (p = 0.2). Therefore, the jellyfish galaxy candidates are not systematically closer to their neighbours than the rest of the galaxies. This result reassures that the mechanism responsible for the jellyfish signatures is most likely RPS rather than tidal interactions or mergers.

At the end of the selection process we reviewed each one of the candidates and applied a flag for possible tidal interactions and/or mergers for galaxies that appear to be too close to a companion. In total, three galaxies were flagged, IDs: 33058, 34033 and 34839. They remain in the ATLAS, but they are not included in the plots and analysis. Throughout the paper we may refer to different groups of JClasses by shortening the nomenclature, e.g. JClasses 3, 4 and 5 to JC345.

\vspace{3mm}

\subsection{Trail Vectors}%ok
Galaxies undergoing RPS often leave trails of gas, dust and recently formed stars behind as they move around the system. Based on these morphological structures it is possible to infer the projected apparent infalling direction of the galaxies \citep{smith10,mcpartland16}. We call this the trail direction of the galaxy and we represent it with a trail vector, this vector should point towards the motion of the galaxy. In this section we describe the method we have followed for assigning the trail vectors as a second stage of the visual inspection.

Each one of the three classifiers independently assigned a trail vector to every jellyfish galaxy candidate as a first stage.
The classification involved two steps: the identification of the most pronounced RPS signature (e.g. tails) and then the recognition of the direction in which this feature is being stripped.
After this stage, the three inspectors reviewed together the individually assigned vectors to yield a final vector with a unanimous agreement. Figure~\ref{fig:hacont} shows some examples of the trail vectors assigned.

\section{Results}\label{sec:res}

\subsection{Morphologies, stellar masses and SED types}
\begin{figure*}
  \begin{minipage}[b]{0.33\linewidth}
    \includegraphics[width=\textwidth]{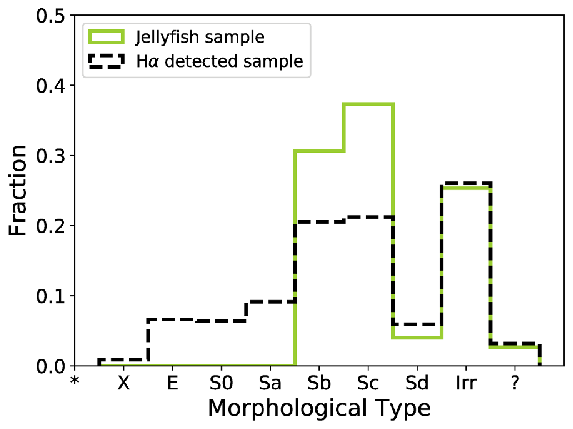}
  \end{minipage}
  \begin{minipage}[b]{0.33\textwidth}
    \includegraphics[width=\textwidth]{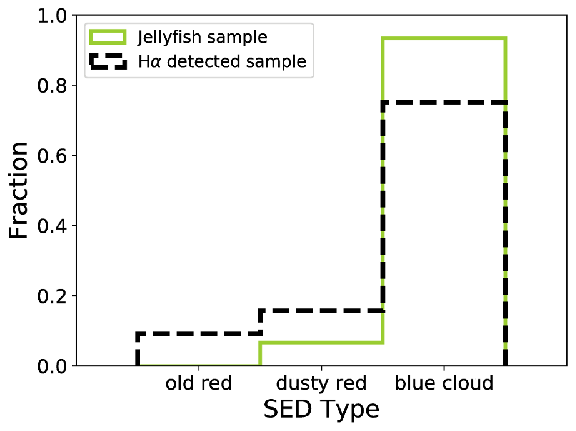}
  \end{minipage}
  \begin{minipage}[b]{0.33\textwidth}
    \includegraphics[width=\textwidth]{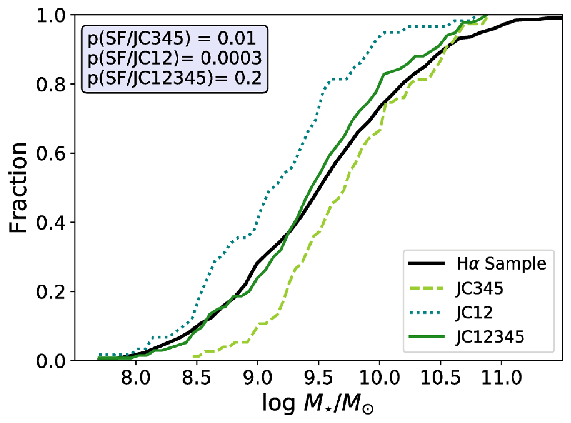}
  \end{minipage}
\caption{Left panel: the histogram of the STAGES morphological types of the jellyfish galaxies compared to the H$\alpha$ sample. Middle panel: the SED types histogram determined by STAGES for both the jellyfish candidates sample and the H$\alpha$ sample. Right panel: cumulative histogram of the stellar mass distributions for the OMEGA-H$\alpha$ sample, the jellyfish candidates sample (JC345), the galaxies with weak RPS evidence (JC21) and all galaxies with JClass higher than 0 (134 galaxies).}
\label{fig:analysis}
\end{figure*}

In this subsection we explore the main properties of our sample of jellyfish galaxy candidates in comparison to the other H$\alpha$ emitting galaxies in the OMEGA sample. We look at morphologies, stellar mass distribution and SED types to find whether the jellyfish phenomenon is associated to galaxies with distinct properties. In Figure~\ref{fig:analysis} we show such comparisons. 

In the left panel of Figure~\ref{fig:analysis} we compare the morphological types assigned by the STAGES collaboration for the galaxies in the whole OMEGA sample and the jellyfish candidates sample. The sample of jellyfish galaxies (JC345) is composed mainly by late-type spirals and irregulars.
In the middle panel of Figure~\ref{fig:analysis}, we show the distribution of SED types for both samples. Based on the SED types of the galaxies, out of the 70 jellyfish galaxy candidates analysed, 66 were found to be part of the blue cloud and 4 as being dusty reds (IDs: 11633, 17155, 19108, 30604).
However, contrary to what could be expected, dusty red galaxies are only a small portion of our sample of jellyfish candidates.
One reason why we may not detect many dusty reds as jellyfish galaxies might be due to the fact that these galaxies, despite having relatively high SFRs (only four times lower than that in blue spirals at fixed mass, \citealt{wolf09}), have significant levels of obscuration by dust which might hamper the identification of the jellyfish signatures.
Another reason for that is that we selected jellyfish galaxy candidates within a parent sample of H$\alpha$ emitting galaxies that had already a low fraction of dusty red galaxies ($\approx$ 15$\%$). As these galaxies have low star formation it is harder to perceive the morphological features of RPS.
Dusty red galaxies have been previously studied in this same system \citep{wolf09} and RPS was suggested to be the main mechanism acting in these galaxies \citep{bosch13}.
While in \citet{bosch13} one of the main evidences suggesting the action of enhanced RPS was the existence of disturbed kinematics without disturbed morphologies, in our study we strongly base our selection on such morphological distortions.
Both our jellyfish galaxy candidates and the dusty red galaxies show different characteristics that can be correlated to the effect of RPS. Nevertheless, they might be tracing different stages of the same phenomenon, where dusty red galaxies have more regular morphologies, but disturbed kinematics. Our sample of morphologically disturbed jellyfish galaxy candidates may be showing the stage where the features of RPS are the most visible and the star formation rates are enhanced.

%are going through a stage of RPS where the gas is being disturbed but does not show morphological disturbances.

Finally, the right panel of Figure~\ref{fig:analysis} shows the stellar mass distribution in a cumulative histogram for the different samples. We can see in the cumulative mass distribution that the jellyfish candidates (JC345) have higher masses than the other galaxies in the OMEGA sample (a KS test returns a pvalue of 0.01).
Nevertheless, for less massive galaxies the visual evidence for gas stripping is less noticeable, specially in the continuum. In this way, RPS events in less massive galaxies may not be selected or may end up being assigned lower JClasses, as 1 or 2, which may cause a bias towards selecting more massive galaxies as jellyfish galaxy candidates.
We check this hypothesis by adding the weaker cases JC12 to the plot, they appear to be less massive than the parent or the jellyfish sample. If we merge all JClasses together, we find that it follows very closely the mass distribution of the parent sample with no statistically significant difference (p = 0.2). Thus, we conclude that the apparent shift towards higher masses in this sample of jellyfish galaxy candidates is due to a selection bias.

\subsection{Environmental Properties}\label{ssec:env}%
\begin{figure}
    \centering
    \includegraphics{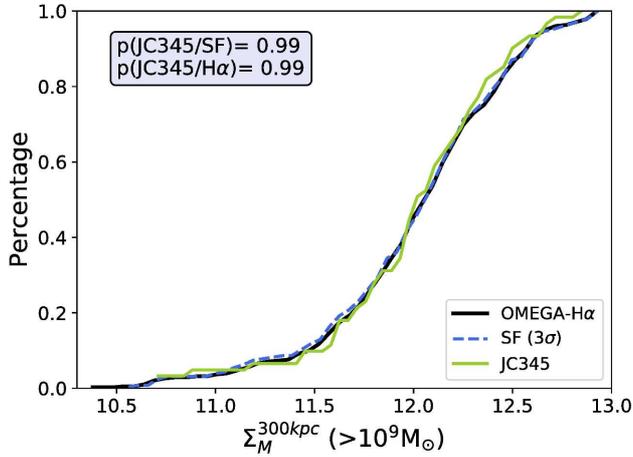}
    \caption{Cumulative histogram of the distribution of the galaxies by stellar matter density. We compare the jellyfish candidates (green solid line) to the OMEGA galaxies with active star formation (blue dashed line) and we plot the OMEGA-H$\alpha$ sample (grey solid line) for reference.}
    \label{plt:envdens}
\end{figure}

To test effects due to environment we have compared the environments where jellyfish candidates and the star-forming galaxies in OMEGA reside. 
We first compare the stellar mass density of both populations. This is calculated as described in \citet{rodriguezdelpino17} and by following the procedure of \citet{wolf09}. We use the $\Sigma_M^{300kpc} (> 10^9 M_{\odot})$ parameter.
Figure~\ref{plt:envdens} shows the cumulative histogram for the OMEGA-H$\alpha$ sample, the OMEGA-SF and the jellyfish candidates (JC345).
We find no significant difference among the samples.
However, it is important to note that our range of environmental densities is not broad and there may exist some behaviour outside of this range that we might not be detecting.

\begin{figure}
\centering
\includegraphics{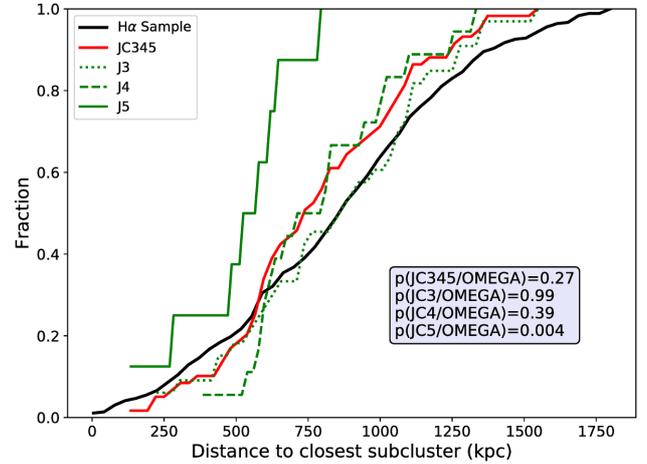}
\caption{Cumulative histogram of the distribution of projected distances from the galaxies to the closest sub-cluster. We compare the jellyfish candidates by JClass to the OMEGA-H$\alpha$ sample. The green lines show the jellyfish candidates distribution by JClass: 5 (solid line), 4 (dashed line) and 3 (dotted line). The OMEGA-H$\alpha$ sample distribution is represented by the black solid line.}
\label{plt:rad}
\end{figure}

We have also checked the relation between the sample and the environment as function of the projected radial distance between the galaxies and the positions of the sub-cluster centres.
Here, in order to avoid the contamination by the galaxies in-between two sub-clusters, we are only analysing the galaxies enclosed in the inner regions of the virial radius $R_{200}$ of each sub-cluster. In case of overlapping, which occurs with A901a and A901b, the galaxies are considered members of the sub-cluster they are closest to.
We find that the whole distribution of jellyfish candidates (JC345) is not significantly different from the OMEGA sample (p=0.2). We have then divided the galaxies in subsamples of different JClasses, which is shown in Figure \ref{plt:rad}.
We find that the higher the JClass, the closest they are to a sub-cluster centre. Performing KS tests in these three distributions we find the following values: p=0.004 for JC5, p=0.4 for JClass 4 and p=0.98 for JClass 3. Such behaviour is therefore only found to be highly significant for the strongest jellyfish candidates.
However, these results are not entirely reliable given the small number of objects in the samples tested.

\subsubsection{Spatial distribution of the ram pressure stripping events}
\begin{figure*}
\includegraphics{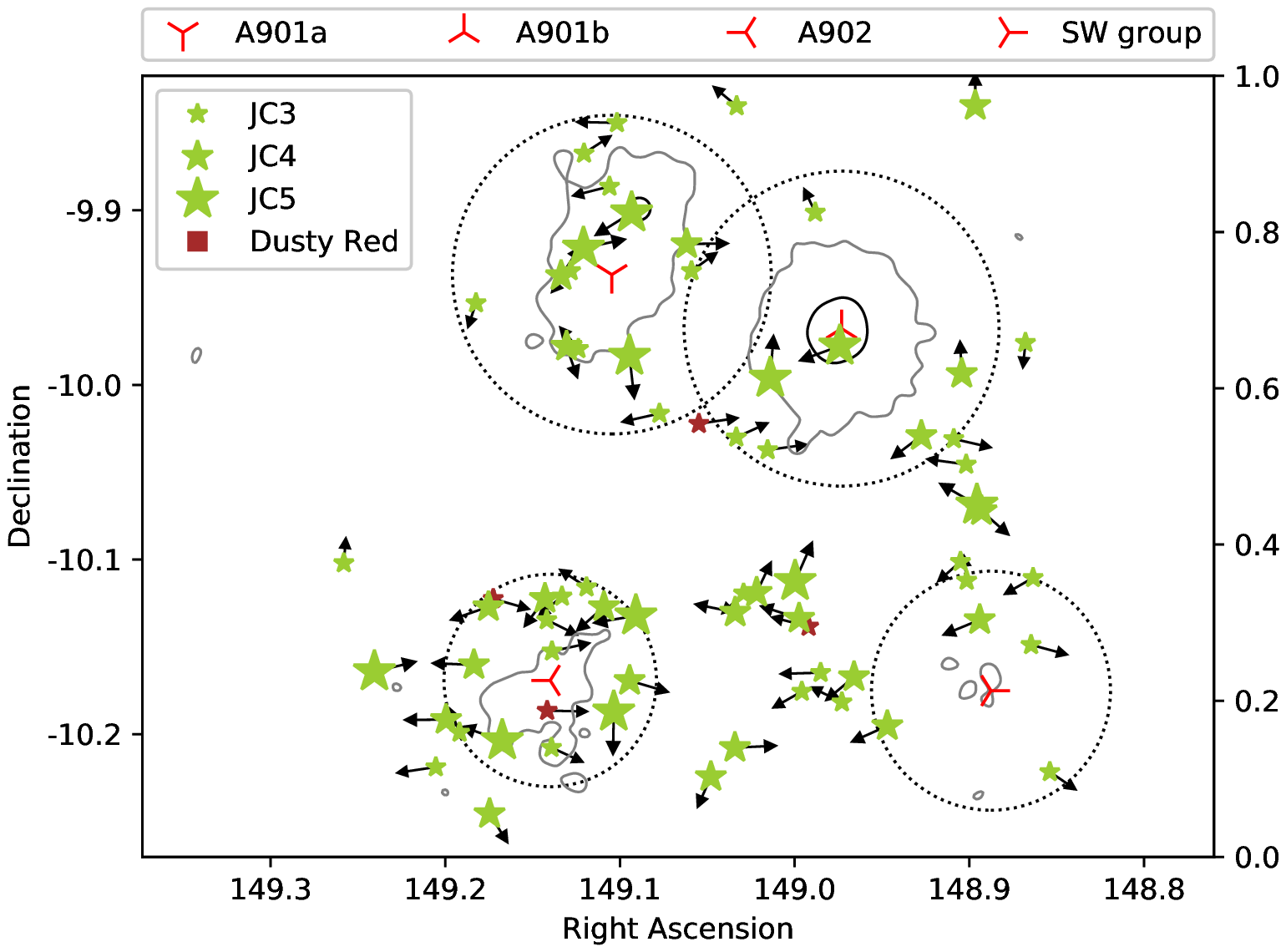}
\caption{Spatial distribution of the jellyfish galaxies around the four sub-clusters in the A901/2 system. Each sub-cluster is labelled and have the circles showing their virial radius $R_{200}$ (dash-dot black circles). The stars represent the jellyfish candidates according to the legend. The dusty red galaxies are marked in red. All of their respective trail vectors are shown as arrows. The grey contours show the gas density as measured from the x-ray emission, they are divided in three levels of significance: 3$\sigma$ (solid dark line) and 2$\sigma$ (solid gray line).}
\label{fig:spatial}
\end{figure*}

In Figure~\ref{fig:spatial} we explore the projected spatial distribution of the candidates on the system.
We also show the contours of the X-ray emission divided into two different levels of significance: the black lines contour a 3$\sigma$ level and the gray lines contour a $2\sigma$ level. The X-ray comes from the emission of the hot gas and traces its distribution. The highest level contour allows us to see where the majority of the hot gas is located and the second contour assists in establishing the extent of its distribution around the system.
We find that approximately 40\% of the galaxies are located outside the virial radius of the sub-clusters. However, for the most massive sub-clusters (A901a and A901b) the jellyfish galaxies are mostly located within the virial radius -- only around 30\% of the galaxies are outside the virial region. Whereas in less massive ones (A902 and SW group) their distribution is more extended -- approximately half of the sample is located outside the virial radius of these sub-clusters. These galaxies are probably not yet attached to the gravitational potential of any of the sub-clusters. If we consider only the most compelling candidates (JC45), we see that half are located in the A902 system, however only two are found inside the virial radius of the SW group.

We also show in Figure \ref{fig:spatial} the respective trail vector of each galaxy. We can infer whether the galaxies appear to be falling towards or moving away from the sub-cluster centres.
For quantifying that, we have calculated the angle between the trail vector and a vector pointing in the direction of the closest sub-cluster centre in projected distance.
If the absolute value of this angular difference is smaller than 90$^{\circ}$ then we say the galaxy is moving towards the system and, if the difference is larger than 90$^{\circ}$, then the galaxy is classified as moving away from the system.

Table~\ref{tab:spdist} contains the number of galaxies either falling towards or outwards any of the systems divided by JClasses.
The spatial analysis of these vectors altogether with the position of galaxies around the system suggests that they have no preferential sub-cluster centres to be falling towards or outwards.
No sub-cluster shows a significant difference between the infalling towards/outwards numbers and as we restrict the analysis to each sub-cluster, however, on these circumstances we are prone to low number statistics.
Our results are in contrast with those found by \citet{smith10} for jellyfish galaxies in the Coma cluster where they are mostly falling towards the cluster centre. An important note is that we are limiting our study to H$\alpha$ emitting galaxies while in \citet{smith10} the sample is limited to UV emitting galaxies covered with GALEX, however, this should not drastically change our findings. Nevertheless, the differences might be due to the fact that the dynamics of A901/2 are much more complex and is a still evolving system, whereas Coma is a more relaxed cluster.

\begin{table}
\centering
\begin{tabular}{|c|c|c|c|c|c|}
\cline{0-5}
\hline
Cluster                   & Direction & JC5 & JC4 & JC3 & Total \\ \hline
\multirow{2}{*}{A901a}    & towards   & 2   & 1   & 2   & 5     \\ \cline{2-6} 
                          & outwards   & 1   & 2   & 7   & 10    \\ \hline
\multirow{2}{*}{A901b}    & towards    & 1   & 1   & 4   & 6     \\ \cline{2-6} 
                          & outwards   & 1   & 2   & 4   & 7     \\ \hline
\multirow{2}{*}{A902}     & towards    & 2   & 3   & 5   & 10    \\ \cline{2-6} 
                          & outwards   & 2   & 8   & 5   & 15    \\ \hline
\multirow{2}{*}{SW group} & towards    & 0   & 2   & 2   & 4     \\ \cline{2-6} 
                          & outwards   & 2   & 3   & 7   & 12    \\ \hline
\cline{0-5}
\end{tabular}
\caption{Distribution of the projected direction of motion of the candidates per sub-cluster and per JClass, as implied by the trail vectors assigned.}
\label{tab:spdist}
\end{table}

Given that the effect of ram pressure depends strongly on the density of the hot gas \citep{gunngott72}, in principle we would expect a correlation between the distribution of the hot gas and the jellyfish galaxies.
In our case, this may explain why there are so few cases of evident jellyfish in the SW group since it is the region with the weakest x-ray emission, thus less hot gas. This also explains why the strongest candidates (JC45) tend to gather in the inner regions of the clusters.
However, for the cases outside the inner regions of the sub-clusters, the influence of the merging system has to be taken into account as well.
The effect of cluster mergers in the observation of RPS events has already been suggested in the Abell 2744 system by \citet{owers12}. Three of the four jellyfish galaxies were found closely to the gradients in the X-ray emission, features of the cluster merging, suggesting that cluster mergers can trigger RPS events.
This phenomenon has also been hinted in the work of \citet{mcpartland16}, where their results suggest that extreme RPS events linked to cluster mergers. The fact that the Abell 901/2 multi-cluster system holds a rich jellyfish population is a compelling evidence that the unrelaxed nature of interacting systems may cause an enhancement of the fraction of jellyfish galaxy events.As well as increasing the number of cases, the distribution of RPS events in merging systems would not only follow the distribution of hot gas but also its dynamics. The RPS phenomenon has a square dependency on the relative velocity between the galaxy and the hot gas, while the dependency is linear with the density of the hot gas \citep{gunngott72}. Interacting clusters provide much greater velocities than single relaxed systems on the frontiers of the interaction. For this reason, it is not unexpected that the jellyfish galaxies would not follow an even distribution around and towards the sub-cluster centres. These galaxies could be actually tracing the regions on where the relative velocity increases dramatically due to the interactions of the sub-clusters. A simulation work on the jellyfish galaxies in the A901/2 system shows the tendency of the galaxies gathering around the regions where the relative velocity of the ICM is higher \citep{ruggiero19}.

\subsubsection{Projected Phase-Space Diagram}
The phase-space analysis for the OMEGA-H$\alpha$ sample has been performed in \citet{weinzirl17}. Among other interesting results, it was found that there is no change in the sSFR of the star-forming galaxies at fixed mass throughout the cluster environment. This suggests that pre-processing of galaxies during the infall is a dominant mechanism in quenching the star formation.

\begin{figure*}
    \centering
    \includegraphics{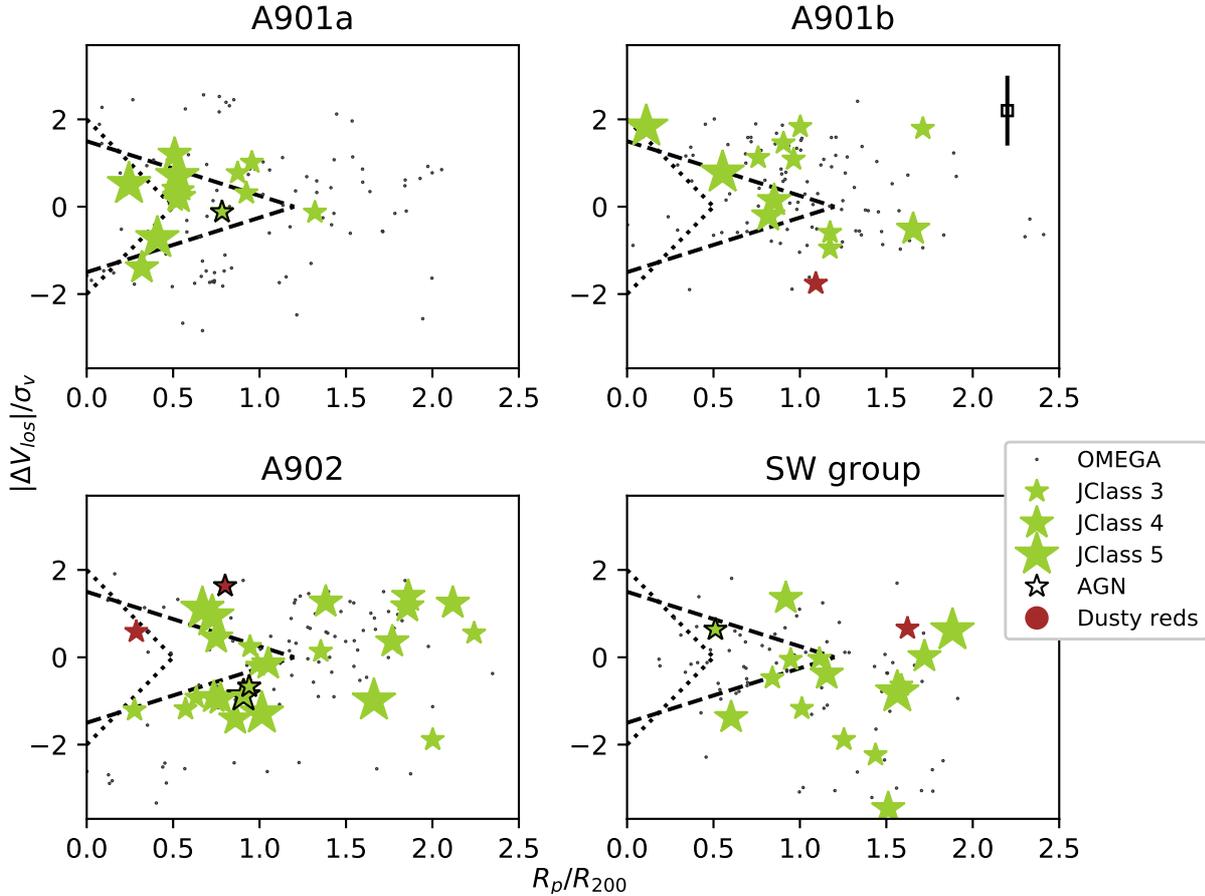}
    \caption{Phase space diagram for the jellyfish galaxy candidates divided by sub-cluster according to the legend. The sample is divided by JClass and represented by star symbols according to the legend. Galaxies within this sample that are hosts to an AGN are represented as a square and the dusty red galaxies are painted in red. The open gray circles in the background represents the OMEGA-H$\alpha$ galaxies that show no morphological evidence of RPS. We analyse two fiducial boundaries: Boundary 1 \citep{jaffe15} and Boundary 2 \citep{weinzirl17}.}
    \label{plt:phsp}
\end{figure*}

In Figure \ref{plt:phsp} we show the most secure jellyfish candidates (JC45) and analyse their location in a projected phase-space diagram for each sub-cluster system.
We separate the galaxies by sub-cluster according to the closest sub-cluster centres in projected angular distance. In this diagram we analyse two fiducial radii, the Boundary1 is defined as $R_p/R_{200} \leq 1.2, |\Delta V_{los}/\sigma_{scl}| \leq 1.5 - 1.5/1.2 \times R_p/R_{200}$ and comes from \citet{jaffe15} which was used for analysing the A963\_1 system that lies at z$\sim$0.2 and is close in mass to Abell 901a.
Boundary 2 is defined by $R_p/R_{200} \leq 0.5 e |\Delta V_{los}/\sigma_{scl}| \leq 2.0 - 2.0/0.5 \times R_p/R_{200}$ and was taken from \citet{weinzirl17} that studies in detail the properties of the OMEGA galaxies in the phase-space diagram. The boundaries have the purpose to trace the frontier of the gravitational influence of the sub-clusters. However, it is important to note that the A901/2 multi-cluster is an unrelaxed system and the use of boundaries in the phase-space diagram analysis should be considered as a rough approximation. $V_{los}$ represents the velocity in the line of sight of the galaxies and $\sigma_{scl}$ represents the velocity dispersion of the sub-cluster. 

The projected phase-space diagram divided by sub-centre complements the information provided in Figure~\ref{plt:rad}.
The strongest cases seem to gather closer to the centre and to the boundary of the virialised regions for the most massive clusters. 

As for their velocities, we find that from JC345 sample, only 27 candidates are at high velocities ($\Delta V_{los}/\sigma_{cls}$ > 1), in which three are JClass 5, eleven are JClass 4 and fourteen are JClass 3.
We notice that our candidates do not show particularly high velocities, however, we are only probing the relative velocity on the line of sight to the sub-clusters. As discussed in Subsection~\ref{ssec:env}, since the A901/2 system is in interaction, the dominant velocity would be in the hot gas motion as the system evolves and we can not estimate that from the projected velocity of the galaxies.

\subsubsection{Missing AGN activity}

We find that out of the 70 jellyfish galaxy candidates, 53 of them are star-forming galaxies and 5 are hosts to an AGN with high probability.
The separation of AGN and star-forming galaxies was done in \citet{rodriguezdelpino17} through a WHAN diagram. We are considering as secure cases only galaxies with a high probability (higher than 3$\sigma$) of belonging to one of these two groups given their nuclear emission.
Our findings suggest that AGN activity is not a strong feature in the sample. Extreme RPS cases have been proposed as a triggering mechanism for AGN activity \citep{poggianti_nat17}. However, the low fraction of AGN hosts in our sample, specially among the JClass 5 galaxies, and their position in the PPS diagram in Figure~\ref{plt:phsp}, points to the scenario that the RPS is not triggering AGN activity in the sample and that the few AGN cases we find do not seem to be correlated to RPS.

We find that no AGN is hosted by a JClass 5 galaxy, only one is hosted by JClass 4 galaxy and the remaining four AGNs are found in JClass 3 galaxies. If we lower the criteria to a 2$\sigma$ probability, we find other 3 less probable cases of AGN activity: one in a JClass 5 galaxy, another in a JClass 4 and the remaining in a JClass 3 galaxy.
Moreover, the most compelling jellyfish candidates (JC45) that are AGN do not seem to fall on the regions where the RPS is expected to be strongest - small radius and high velocities. Both of them are found at larger radii (r > 0.5$ R_p/R_{200}$) and only one is in the high velocity region ($\Delta V_{los}/\sigma_{cls}$ > 1).

Interestingly, even though the AGN activity does not seem related to the RPS, the AGN hosts seem to have relatively higher masses than the rest of the jellyfish candidates sample (4 of them are more massive than $10^{10.2} M_{\odot}$). It may be an evidence that the AGN found in the sample may be more related to the masses of the host galaxies and that the RPS signatures may be a coincidence instead of a trigger. However, the statistics is too low for a definite answer.

\begin{figure*}
    \centering
    \includegraphics{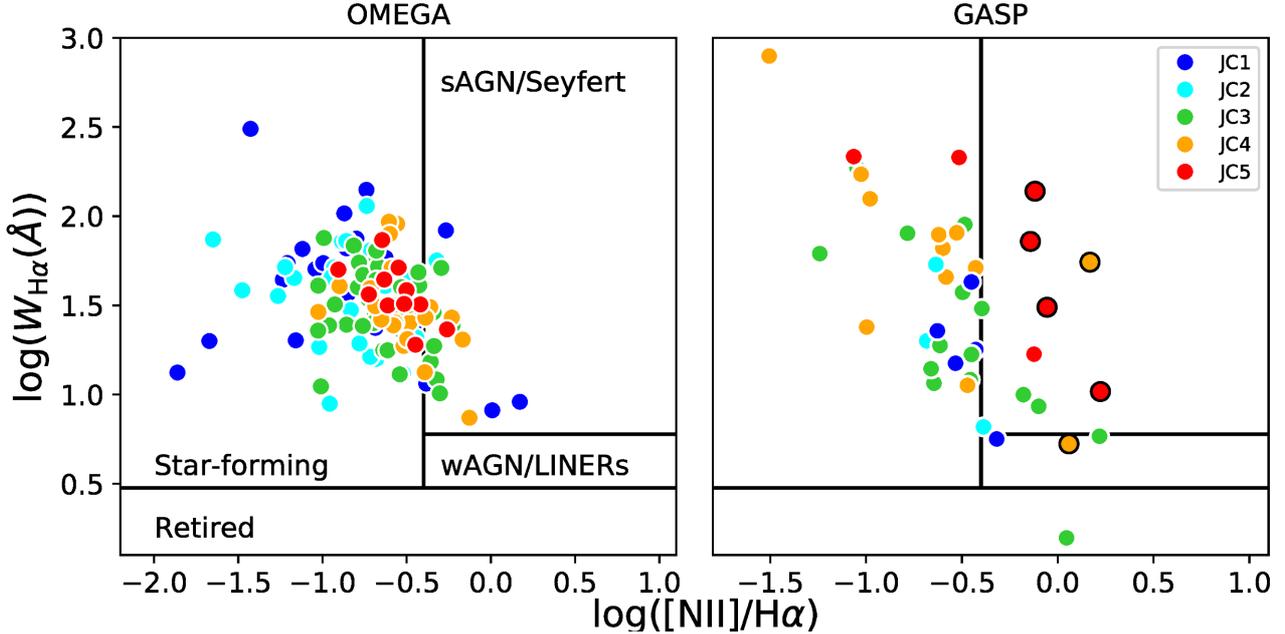}
    \caption{The WHAN diagram for jellyfish galaxies in the public GASP sample (left panel) and for the OMEGA jellyfish galaxy candidates (right panel). Different JClasses are shown in different colours according to the legend. The markers with a black edge are the galaxies present in \citet{poggianti_nat17}.}
    \label{fig:whan}
\end{figure*}

Finally, we have downloaded the publicly available GASP data for 42 jellyfish galaxies. For each MUSE data cube we have selected the integrated spectra in the 6$\times$6 spaxels around the centre of the galaxies, fitted the emission lines and measured, in a similar way to the OMEGA data, EWs and line ratios. We show in Figure~\ref{fig:whan} the WHAN diagram \citep{cid10} comparison of our findings with that of the public GASP sample of jellyfish galaxies. As in \citet{rodriguezdelpino17} we employ the vertical line separation of [NII]/H$\alpha$=0.4 proposed by \citet{stasinska06}. For the sake of comparison we add the JClasses 1 and 2 in this plot as the GASP sample keeps these objects. The trend we find in the OMEGA sample is consistent with what we find in the GASP sample. The majority of galaxies shows ongoing star formation not associated with nuclear activity. We have also generated the BPT diagrams for the GASP sample where this trend is perhaps even more visible. We chose, however, to only show the WHAN diagram as we can compare with the OMEGA jellyfish galaxy candidates sample as well.

\subsection{Star Formation Properties}
\subsubsection{Spatially Resolved Star Formation}
\begin{figure*}
    \centering
    \includegraphics{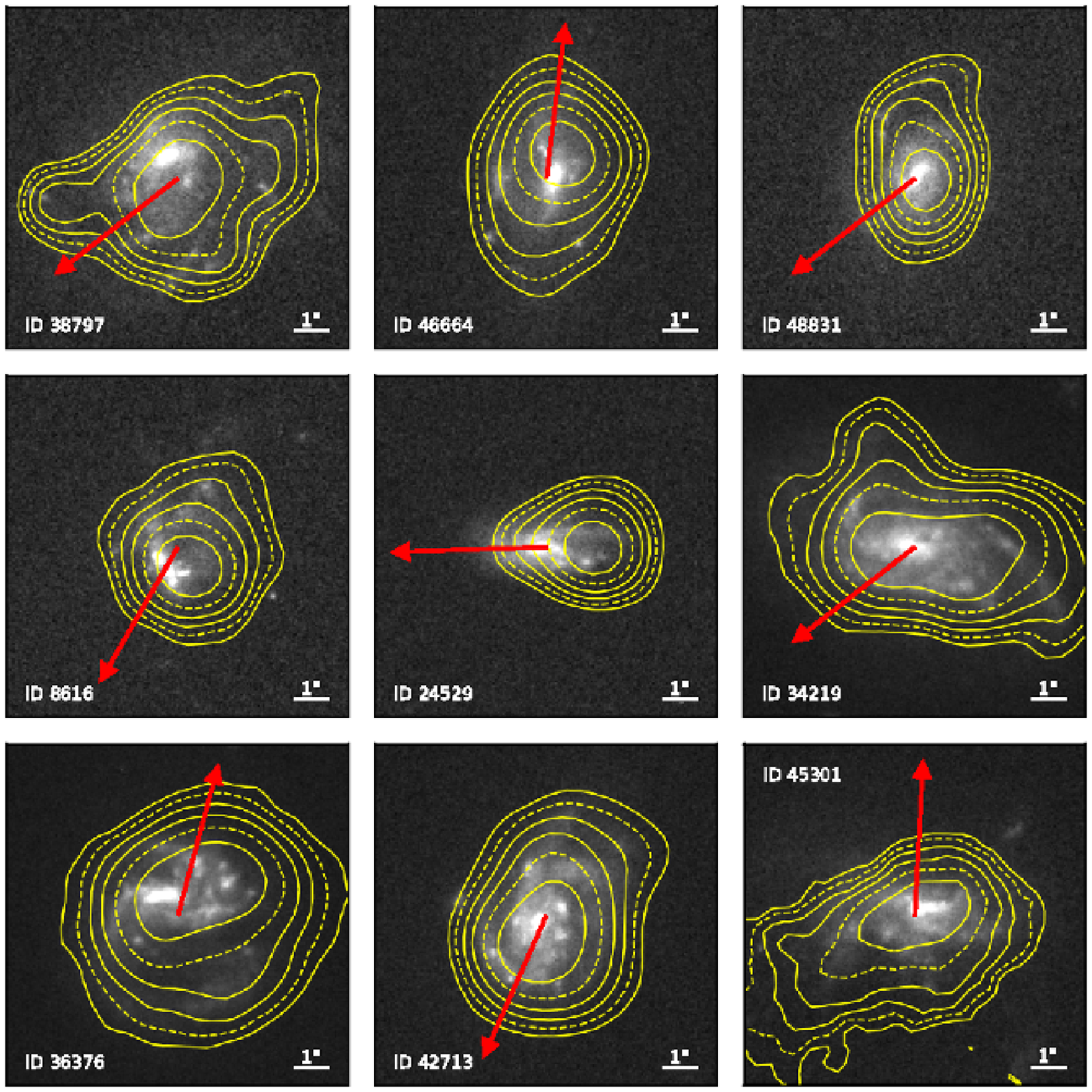}
    \caption{Examples of H$\alpha$ contours and final trail vectors. Top row -- JClass 3; middle row -- JClass 4; bottom row -- JClass 5.}
    \label{fig:hacont}
\end{figure*}

We have studied the H$\alpha$ emission for the jellyfish candidates by analysing the H$\alpha$ emission contours on top of the HST continuum images. The maps generated for the jellyfish candidates and for the other galaxies in the OMEGA sample will further be available and studied in detail in Rodr\'iguez del Pino et al. in prep.. We show some examples in Figure \ref{fig:hacont} for jellyfish galaxy candidates of JClasses 5, 4 and 3.

We show the H$\alpha$ emission contours on top of the HST continuum images together with the final trail vector for all galaxies in the ATLAS. The contours are missing for some galaxies as there were not enough images in the OMEGA continuum and/or around the H$\alpha$ line to build them accurately.
The spatial distribution of the H$\alpha$ emission, for part of the sample, is evidently disturbed and extended and, in some cases, the extension agrees with the trail vector previously assigned. This points towards a scenario that as well as stripping gas out of the galaxy, ram pressure may also enhance star formation activity, both inside and outside the galaxies. The fact that the H$\alpha$ emission is disturbed and extended indicates that the star formation is also taking place where the gas is being stripped out of the galaxy and building the asymmetrical structures we observe.

\begin{figure*}
\includegraphics{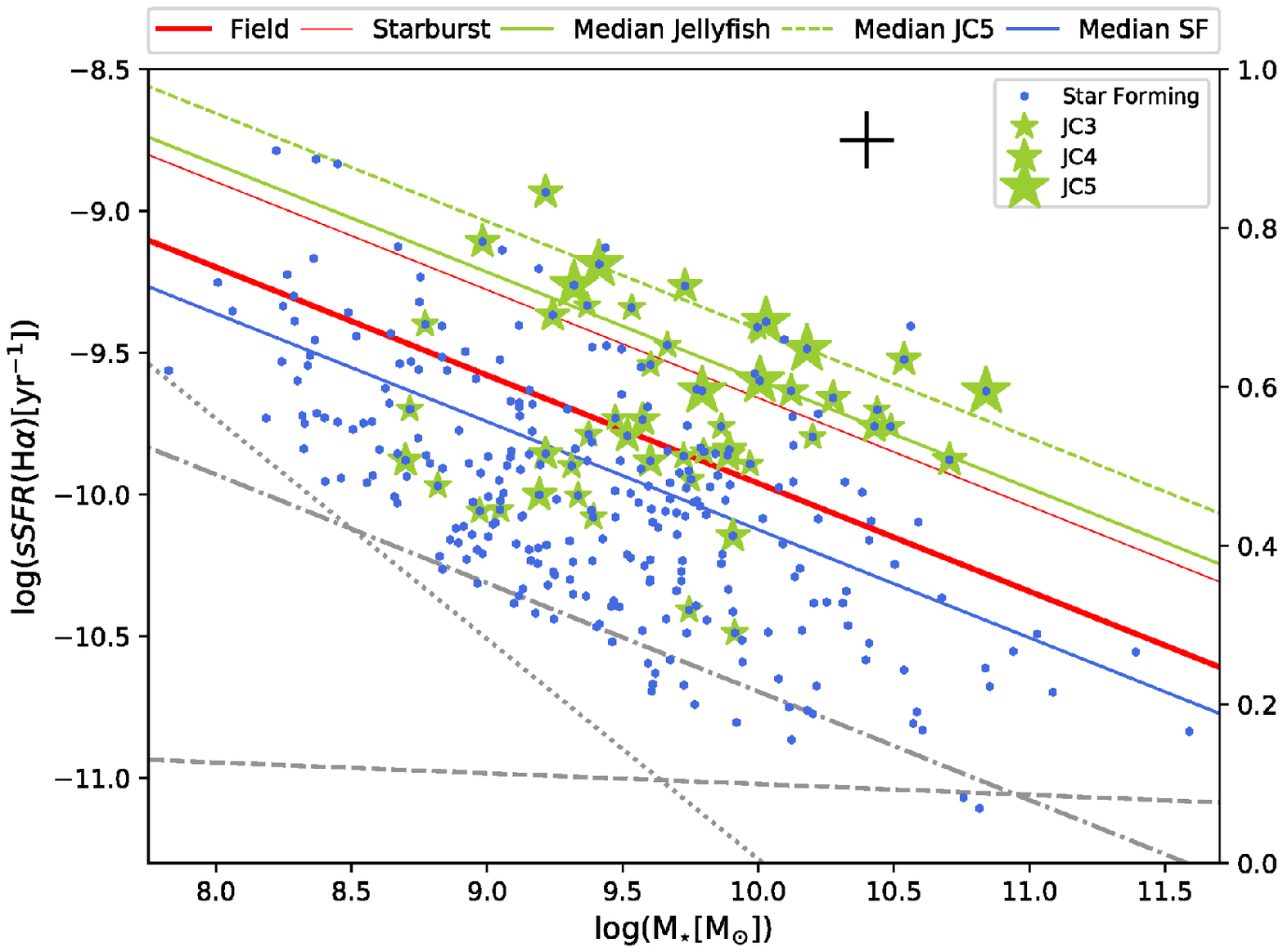}
\caption{Specific star formation rate versus mass: jellyfish galaxies are separated by JClass and represented by the green star symbols, the larger the star the more evident the "jellyfish" morphology. The blue dots represent the OMEGA-SF. The main sequence sSFR-stellar mass relation for the SDSS field galaxies is represented by the red line. The green and blue lines are, respectively, lines that go through the median of the jellyfish (solid for JC345 and dashed for JC5) and star-forming populations with the same slope as the red line. The thinner red line marks a sSFR that is twice that of the main sequence, which has been used to outline starbursts \citep{elbaz11}. The grey lines show the detection limits of the OMEGA survey: H$\alpha$ flux (dotted), equivalent width (dashed) and the lower boundary for the region free from incompleteness (dash-dot).}
\label{fig:ssfr}
\end{figure*}

\subsubsection{Integrated Star Formation}
As for the integrated star formation properties of the candidates, we generate a specific star formation rate (sSFR) versus mass diagram as in \citet{rodriguezdelpino17}, shown in Figure \ref{fig:ssfr}.
We compare the sSFR of the jellyfish galaxy candidates, divided by JClass, to the star-forming galaxies in the OMEGA sample.
We also include in the figure the main sequence of star formation at the same redshift derived from the SDSS \citep[DR7]{abazajian09} field galaxies.
We then draw two more lines with the same slope that goes through the median of each population of galaxies: the green solid line for jellyfish galaxies (JC345) and blue for the star-forming OMEGA galaxies. We find that our jellyfish galaxy candidates sample have higher sSFR than it would be expected for galaxies similar in mass in a field environment. Given that many galaxies in the parent sample have reduced their star formation activity, as seen in \citet{rodriguezdelpino17}, it is striking that most of the jellyfish galaxies are going against this trend and are located above the field relation. In fact, 55$\%$ of the jellyfish galaxy candidates are above the main sequence line.
The process that the jellyfish galaxies are undergoing is producing an enhancement in their star formation activity that places them above the field relation. This happens despite the environmental quenching that is reducing the star formation in the other star-forming galaxies in OMEGA \citep{rodriguezdelpino17}.
To quantify the difference in specific star formation rates, we run a KS test in a cumulative histogram of the sSFR of both populations. The results show that none of the subsamples (JC345, JC45 and JC5) can be part of the same parent population.
Whilst these galaxies have been selected only by visual evidence of RPS it suggests that such mechanism is indeed enhancing the star formation of some of these galaxies.
In Figure~\ref{fig:ssfr}, we also draw a thinner second red line which stands for a sSFR of twice the value of the main sequence, we use it as a lower limit for what we can consider to be starburst galaxies \citep{elbaz11}.
Using this line as reference, 19 of the 70 jellyfish candidates found seem to be undergoing a starburst period. This line has also been used in the work on the Abell 2744 system with 4 jellyfish galaxies where 1 of them showed to be starburst by this definition \citep{rawle14}.
From these 19 starburst galaxies, when separating by JClasses, the starburst phenomenon seems to be correlated with how evident the jellyfish morphology is, where: 8 of the 11 JClass 5 galaxies, 6 of the 22 JClass 4 galaxies and only 5 of the 37 JClass 3 galaxies appear to be starbursts.
An enhancement in the specific star formation rate in jellyfish galaxies has already been suggested by \citet{rawle14}, for only 4 jellyfish galaxies in a merger system, and \citet{poggianti16}, for 344 candidates scattered in several different clusters. Moreover, \citet{vulcani18} find that stripping galaxies show a systematic enhancement in the SFR-mass relation when compared to undisturbed galaxies. However, this is the first time that this effect is observed in a large number of objects in a single multi-cluster system.
This could be explained by thinking about jellyfish galaxies as a quick transition morphology that links different stages of galaxy evolution.
It may be that galaxies undergoing RPS suffer an enhancement in the star formation, specially in the outskirt regions, leading to a starburst episode. This stage soon runs out of available gas as it is being stripped away and then further leads to the quenching of the galaxy.
This transformation could be strongly correlated with the visual features we observe and, as a consequence, correlated with the JClasses assigned: visually more evident phenomena could be marking the phase of the triggering of star-forming, whereas less evident phenomena could be either the pre-SF-trigger or the post-SF-trigger period.

\section{Summary and Conclusions}\label{conc}
In this work we have conducted a systematic search for galaxies that show morphological evidences of gas stripping in the Abell 901/2 system, at z $\sim$ 0.65, and a detailed analysis of their overall properties as part of the OMEGA survey.
The search was conducted over the OMEGA parent sample of 439 H$\alpha$-emitting galaxies.
The final sample is composed by 73 galaxies, classified in 5 different categories of visual magnitudes of the phenomenon named JClasses -- 1 being the weakest evidence of RPS and 5 being the strongest. This is the largest sample of jellyfish galaxy candidates in a single system to date. We flag down 3 galaxies as possible tidal interactions and run the analysis on the remaining 70, in which our main findings are:

\begin{enumerate}

    \item The typical morphologies of the jellyfish galaxy candidates are late-type spirals or irregulars. The sample is dominated by blue cloud galaxies with only 4 being previously assigned a dusty red classification. We have found only 5 AGN host galaxies. Moreover, the jellyfish galaxy candidates appear to be slightly more massive than the other galaxies, which we associate to a visual selection bias.

    \item The jellyfish galaxy candidates spatial distribution and apparent motion around the multi-cluster system does not show an obvious pattern. We find little correlation between the distribution of jellyfish galaxies and hot gas traced by X-ray emission. However, the most evident candidates (JC5) seem to be located closer to the centres of the sub-clusters when compared to the other less evident cases. The two most massive sub-clusters (A901a and A901b) have a larger and more concentrated population of jellyfish galaxies around them, While half of the compelling cases (JC45) are gathered around the intermediate mass system (A902). In fact, the sub-cluster with the lowest mass (SW group) has only two compelling jellyfish candidates (JC45) within its virialised region.
    
    \item We find that the jellyfish galaxy candidates specific star formation rates are higher than the typical main sequence values, despite what happens to the other star-forming galaxies in the system that show significantly reduced star formation rates. In fact, the median trend for the sample shows higher sSFR than the lower limit of the starburst definition we have used from \citet{elbaz11}. Furthermore, we also find evidence of extended and disturbed star formation for part of the sample.

\end{enumerate}

Our interpretation is that the low fraction of dusty reds in the sample of jellyfish galaxy candidates -- 4 out of 70 -- suggests that the galaxies selected through visual evidence are at a later stage of the RPS event than those that only show disturbed kinematics. At first only the gas is affected and the RPS does not significantly impact the morphology of the galaxy. However, the disturbed gas triggers extended star formation that leads to a disturbed jellyfish morphology.
We also find no link between our most compelling jellyfish candidates and AGN activity. Due to the low fraction of AGN within our sample -- 5 out of 70 -- and the fact that the few ones we find are not located in the region of the phase-space diagram where RPS is at its peak, we are not able to link both of these phenomena in A901/2.

%Explain environment
The large number of jellyfish galaxy candidates found is a compelling evidence that RPS events might be enhanced in interacting systems, making multi-cluster systems ideal environments to search for other jellyfish galaxy candidates.
Also, the apparent lack of pattern in the motion and spatial distribution of the sample of candidates around A901/2 might be evidence of how the RPS phenomenon occurs in multi-cluster systems. Since there is added dynamics to the ICM due to the motion of the subclusters, the relative velocity between the galaxy and the hot gas dominates over the factor of the ICM density. Therefore, the distribution and motion of the galaxies do not necessarily follow the hot gas traced by the X-rays.

%Explain star formation
Our findings also point to the enhancement of star formation as consequence of the RPS phenomenon. In our sample of jellyfish galaxy candidates we found a strong correlation between the morphological asymmetry, traced by the JClasses, and high specific star formation rates. This result supports the evolutionary scenario proposed that: at first, the disturbances are only dominant in the gas and star formation is not enhanced; at a later stage, the perturbations work as a trigger of star formation on the outskirt regions of the galaxy creating the morphological features that we identified in this work. The extended star formation enhances the overall sSFR of the galaxy and can cause a starburst period that is probably short lived as the gas continues to be stripped to further cause a quenching in the star formation.

\section*{Acknowledgements}%
This work is based on observations acquired through ESO large Programme ESO188.A-2002 at the \textit{Gran Telescopio Canarias}, installed at the \textit{Observatorio del Roque de los Muchachos} of the \textit{Instituto de Astrof\'isica de Canarias}, on the island of La Palma. We also use observations collected at the European Organization for Astronomical Research in the Southern Hemisphere under ESO program 196.B-0578. This study was financed in part by the \textit{Coordena\c{c}\~{a}o de Aperfei\c{c}oamento de Pessoal de N\'ivel Superior - Brasil} (CAPES) - Finance Code 001. ACS and FRO acknowledge funding from the brazilian agencies \textit {Conselho Nacional de Desenvolvimento Cient\'ifico e Tecnol\'ogico} (CNPq) and the Rio Grande do Sul Research Foundation (FAPERGS) through grants PIBIC-CNPq, CNPq-403580/2016-1, CNPq-310845/2015-7, PqG/FAPERGS-17/2551-0001, PROBIC/FAPERGS and L'Or\'eal UNESCO ABC \textit{Para Mulheres na Ci\^encia}. BRP acknowledges financial support from the Spanish Ministry of Economy and Competitiveness through grants ESP2015-68964 and ESP2017-83197. At last, we are grateful for the valuable comments from the anonymous referee.

%%%%%%%%%%%%%%%%%%%%%%%%%%%%%%%%%%%%%%%%%%%%%%%%%%



\bibliographystyle{mnras}
\bibliography{refer} % if your bibtex file is called example.bib

%\begin{thebibliography}{99}
%\end{thebibliography}

%%%%%%%%%%%%%%%%%%%%%%%%%%%%%%%%%%%%%%%%%%%%%%%%%%

%%%%%%%%%%%%%%%%% APPENDICES %%%%%%%%%%%%%%%%%%%%%

%\appendix

%\section{ATLAS}

%%%%%%%%%%%%%%%%%%%%%%%%%%%%%%%%%%%%%%%%%%%%%%%%%%

% Don't change these lines
\bsp	% typesetting comment
\label{lastpage}
\end{document}